\documentstyle[aps,prl,epsfig,twocolumn]{revtex}

\draft

\begin{document}

\title{ Interference of Bose-Einstein condensates in momentum space}
\author{ L. Pitaevskii$^{a,b}$ and S. Stringari$^a$}
\address{$^a$Dipartimento  di Fisica, Universit\`{a} di Trento,}
\address{and Istituto Nazionale per la Fisica della Materia, 
I-38050 Povo, Italy}
\address{$^b$Kapitza Institue for Physical Problems, 117334 Moscow, Russia}

\date{\today}

\maketitle

\begin{abstract}
We suggest an experiment to investigate the linear superposition of  two 
spatially separated Bose-Einstein condensates.
Due to the coherent combination
of the two wave functions, 
the  dynamic structure factor, measurable 
through inelastic photon scattering at high momentum transfer $q$, is predicted
to  exhibit 
interference fringes with frequency period $\Delta\nu = q/md$
where  $d$ is the distance
between the condensates.
We show  that the coherent configuration 
corresponds to an eigenstate
 of the physical  observable measured in the  experiment
and   that the relative phase of the condensates is 
hence created through the measurement process.

 \end{abstract}


\narrowtext

Dilute Bose-Einstein condensed gases behave like classical matter waves.
This striking feature has been 
directly confirmed by several recent 
experiments  \cite{mit,jila,kasevich,nist}. In 
particular,  very clean interference patterns 
generated by two overlapping condensates have been 
observed through absorption imaging techniques \cite{mit}. 
Interference phenomena produced by matter waves are key features
underlying the quantum 
mechanical behaviour of matter, so it is of considerable interest to understand
what is the new  role played by Bose-Einstein condensation. 

Bose-Einstein condensed gases  can be regarded as classical objects
 because, according to
Bogoliubov prescription, the corresponding field operator 
can be replaced by a classical field, resembling the classical limit
of quantum electrodynamics. 
Differently from the case of the electromagnetic field
which is governed by the  Maxwell equations, 
the field associated with
a Bose-Einstein condensate obeys equations of quantum nature
which reduce, for dilute and cold gases, 
to the
Gross-Pitaevskii equation \cite{GP}
\begin{equation}
i\hbar{\partial\over \partial t} \Psi = (-{\hbar^2 \nabla^2 \over 2m} + V_{ext} + 
g\mid\Psi\mid^2) \Psi
\label{GP}
\end{equation}
where $V_{ext}$ is the external potential  confining the
gas, and
$g =4\pi\hbar^2a/m$ is the interaction coupling constant, 
fixed by the $s$-wave scattering length $a$.
The field $\Psi$ has the meaning of an order
parameter and is often called the "wave function of the condensate".
As a consequence of the 
quantum nature of equation (\ref{GP}), key
features of the  field $\Psi$, like for example 
interference patterns, depend
explicitly on the value of the Planck constant.

The fact that two overlapping condensates behave as coherent matter waves 
giving rise to interference is not however 
obvious. In a  similar context  Anderson \cite{anderson}
raised  the intriguing
question "do two superfluids which have never "seen" one another possess a
definitive phase?". This question has been the object of
theoretical speculations  \cite{leggett} and has been
more recently
reconsidered 
\cite{jav,wallis,burnett,walls,castin} 
after the experimental realization of BEC in trapped atomic gases \cite{bec}.
The point of view shared by 
most authors 
is that the relative phase 
between two condensates is "created" during the 
measurement. In other words, even if the initial configuration is not 
 coherent and the relative phase between 
the condensates is not fixed, one can still observe interference in a single
realization of the experiment. This opens new interesting
perspectives  in the field of quantum measurement in macroscopic 
systems.
However, measuring fringe patterns in the density profiles
requires  overlapping of the condensates in coordinate space and 
one cannot exclude the possibility that interactions among atoms block  
the relative phase before measurement \cite{burnett,rokshar}. 

In this paper we propose an alternative way to investigate interference and
coherence effects, by exploring  the behaviour of the condensate 
in {\it momentum}
rather than in {\it coordinate} space. 
Our proposal is stimulated by the
recent experiment of \cite{bragg} where, by measuring the dynamic 
structure factor at high momentum transfer via stimulated
two-photon Bragg scattering, it was possible to prove  that a single condensate
does not exhibit  phase fluctuations, i.e "it does not consist of
smaller quasi-condensates with random relative phase" \cite{bragg}.  
 In fig. 1
we show a useful  
configuration  where two parallel  trapped condensates are 
located at distance $d$ along the $x$ axis and are described 
by the
order parameters $\Psi_a$ and $\Psi_b$. 
The condensates have no overlap in coordinate space but
 can exhibit  interference in momentum space.
Geometries of this type are now becoming available via
optical confinement techniques. 
In the presence of coherence the
order parameter of the whole system is given by the linear combination 
$\Psi_c = \Psi_a + e^{i\phi}\Psi_b$.  
The corresponding many-body function 
is  $(c^{\dagger})^N\mid >$ where $c^{\dagger}$ is
the particle creation operator in the  state $\Psi_c$ and
$\mid >$ is the vacuum of particles.  
In the following  we assume, for simplicity,
that the two  condensates have the same shape, contain 
the same average  number of atoms ($N_a=N_b=N/2$) and
that the potential separating the two condensates is large enough to exclude 
 overlap between the two wave functions. Under these conditions we can write
$\Psi_a({\bf r})= \Psi_0({\bf r}+{\bf d}/2)$
 and $\Psi_b({\bf r})= \Psi_0({\bf r}-{\bf d}/2)$ where $\Psi_0$ is the 
order parameter of a single condensate 
obeying the Gross-Pitaevskii equation (\ref{GP}) 
and normalized to $\int d{\bf r} \mid \Psi_0\mid^2=N/2$. 
In momentum space the order parameter takes the  form
\begin{equation}
\Psi({\bf p}) = e^{-ip_xd/2\hbar}\Psi_0({\bf p})+
e^{i(\phi + p_xd/2\hbar)}\Psi_0({\bf p})
\label{Psip}
\end{equation}
where $\Psi_0({\bf p})=(2\pi\hbar)^{-3/2}\int d{\bf r}e(-i{\bf p \cdot r}/\hbar)
\Psi_0({\bf r})$. .

Let us calculate 
the average value of the momentum distribution operator 
${\hat n}({\bf p}) = {\hat \Psi}^{\dagger}({\bf
p}){\hat \Psi}({\bf p})$,  where   ${\hat \Psi}({\bf p})$ is the field operator in momentum representation. 
For the coherent configuration (\ref{Psip}) one finds the result
$n({\bf p}) = < {\hat n}({\bf p})>=2(1+\cos(p_xd/\hbar+\phi))n_0({\bf p})$
which exhibits interference fringes  
with period $\Delta p_x = 2\pi\hbar/d$ \cite{note}. In this equation 
$n_0({\bf p}) = \mid\Psi_0({\bf p})\mid^2$ is the momentum distribution
of each condensate. 

In the absence of coherence the many-body wave function has instead the form
$(a^{\dagger})^{N_a}(b^{\dagger})^{N_b}\mid >$ where $a^{\dagger}$ and 
$b^{\dagger}$ are the particle  creation operators in the 
states $\Psi_a$ and $\Psi_b$ respectively. 
This state corresponds to two independent
condensates with fixed number of atoms $N_a$ and $N_b$  respectively,
and the average  of  the momentum distribution operator takes the value
$n({\bf p})=2n_0({\bf p})$ which, as expected, does not exhibit interference.

The momentum distribution of the condensate 
can be investigated experimentally  by
measuring the dynamic structure factor at high energy and momentum transfer. 
 A first useful description is provided 
by impulse approximation  \cite{IA}
\begin{equation}
S(q,E)=\frac{m}{q} \int n(Y,p_y,p_z)dp_ydp_z
\label{Sq}
\end{equation} 
which relates the dynamic structure factor to
the so called longitudinal momentum distribution 
$\nu(p_x)= \int n(p_x,p_y,p_z)dp_ydp_z$. In (\ref{Sq}) $E$ and
$ {\bf q}$
 are the energy and momentum transferred by the photon
to the system and 
 $Y={m\over q}(E-\frac{q^2}{2m})$
is the relevant scaling variable of the problem \cite{Y}.
The vector $\bf q$ has been taken along the $x$ direction. 
A remarkable feature of 
impulse approximation is that 
the quantity $qS(q,E)$ depends on $q$ and $E$
only through the scaling variable $Y$.
Impulse approximation assumes that the system,
after  scattering with the photon, can be described in terms of a scattered 
atom propagating with momentum ${\bf p}+{\bf q}$ and $(N-1)$ atoms 
remaining in the unperturbed configuration.
This approximation has been extensively used  to
analyze the momentum distribution of varius classical and quantum 
systems, including 
liquids and solids through deep inelastic neutron scattering.
In particular it has been employed to extract the condensate fraction 
of superfluid $^4He$ \cite{sokol}.
The impulse approximation is  accurate if one can ignore final state
interaction effects which are responsible for both a shift of the 
peak energy with respect to the free recoil value $ E_r= q^2/2m$ 
and for a broadening of the function $S(q,E)$. First
measurements and theoretical estimates
of these effects in a cold trapped Bose gas have been presented
in \cite{bragg}. For   momentum transfers significantly smaller than the inverse
of the scattering length one can safely use
Bogoliubov theory. 
The relative shift of the peak is a small effect, 
fixed by 
the ratio $\mu/E_r$ where $\mu=gn(0)$ is the chemical potential 
and $n(0)$ is the central density of the gas. Typical values 
in the experiment  of \cite{bragg} 
are  $E_r\sim 20-100 \mu$, depending on the density of the 
sample. The 
broadening due to interactions, of the order of the chemical potential, should
be compared with the Doppler broadening 
$\Delta E_{Doppler} \sim \hbar q/mR_x$,
 fixed by the width $\hbar/R_x$ of the momentum ditribution of the
 condensate and increasing linearly with $q$. Here 
$R_x$ is the radius
of the condensate in the $x$ direction, determined 
by the Thomas-Fermi relation
$\mu=m\omega_x^2R^2_x/2$,  where $\omega_x$ is the radial frequency of  
the  harmonic
potential trapping each condensate. 
The condition derived in \cite{bragg} for the Doppler effect  being
the leading effect can be put in the form
$E_r/ \mu > 0.02 (\mu / \hbar \omega_x)^2$.
Since the ratio $\mu/\hbar\omega_x$ is  large for high density samples,
this condition  can become rather severe. For example, in the
experiment of \cite{bragg} it is well satisfied only for the low
density samples. 
   
A striking feature predicted by (\ref{Psip}) is that, in 
the presence of coherence, also the dynamic structure factor   exhibits
interference. In fact, inserting the corresponding result for the
momentum distribution into (\ref{Sq}), one finds
\begin{equation}
S(q,E)=2[1+\cos(Yd/\hbar+\phi) ]S_0(q,E),
\label{Sinterf}
\end{equation}
where $S_0(q,E) $ 
is the dynamic form factor relative to each condensate. The fringes 
have a period
\begin{equation}
\Delta\nu = {\Delta E \over 2\pi \hbar} =  {q\over m d}
\label{lambdap}
\end{equation}
 and their position is
fixed by the value of the relative 
phase $\phi$ which can be consequently measured.
Notice that the ratio between the Doppler width of $S(q,E)$ and
the distance between two fringes scales as $\sim d/R_x$. A smoothing of the 
interference signal is  expected to occur if one includes 
 final state interaction effects. 
It is worth noticing that 
result (\ref{Sinterf}) holds only if the
energy  of the  scattered atoms, of the 
order of the recoil energy $\sim q^2/2m$, is 
larger than the height of the barrier separating  the two condensates
\cite{note2}. In fact
only in this case will the reflection of the scattered atom from the barrier
be negligible and  the 
dynamic structure factor $S(q,E)$ be 
sensitive to the  phase
$\phi$ of the two condensates. 
Actually
the scattered atoms provide the physical coupling  between
the two condensates.
In the absence of coherence, 
or if the recoil energy is smaller
than the height of the barrier, one has instead $S(q,E) = 2 S_0(q,E)$.   
In fig.2 we show a typical prediction for the dynamic structure function 
in the absence (dashed line) and in the presence of coherence (full line) 
between the two condensates. 
The curves have been calculated for a gas
of sodium atoms trapped in two identical cigar shaped 
harmonic traps with frequencies $\nu_{z}=10Hz$, $\nu_{\perp}=100Hz$ and central
density equal to  $0.5\times 10^{14}cm^{-3}$.
The distance between the centers of the trap is  $d=4R_x \sim 33 \mu m$ 
and the momentum transfer $q$ is equal to $21.3\hbar (\mu m)^{-1}$. With
such parameters the broadening of the
dynamic structure factor due to final state interactions
can be shown to be smaller than 
the period
of interference fringes. The visibility of fringes would  become
even better  
by increasing the momentum transfer $q$ and/or reducing the
density of the sample.

It is interesting to compare the interference fringes 
exhibited by the momentum distribution with the ones characterizing
the density of two expanding and overlapping condensates \cite{mit}. 
A simple estimate
 of the density fringes  
is obtained by neglecting interaction effects
between the two condensates during the expansion. 
This is a good approximation if the two condensates are initially well separated
in space.
 Let us write the order parameter
of each expanding condensate in the form $\Psi_0=
\sqrt{n_0}e^{i\chi}$ where $n_0$ is the density and $\chi$ is the phase.
The interference patterns of  the density $n({\bf r})
=\mid\Psi({\bf r})\mid^2$ are fixed by the phase difference 
$\chi({\bf r} +{\bf d}/2)
- \chi({\bf r} -{\bf d}/2)$. In the geometry of fig. 1 
this difference approaches very rapidly the 
asymptotic value
 $mxd/\hbar t$ corresponding to a relative velocity
$v=d/t$ of the two condensates in the $x$ direction
\cite{dum,rmp}.  In the presence of coherence the density profile 
of the overlapping clouds  takes the form
$n({\bf r},t) =  n_a +   n_b
+ 2 \sqrt{n_an_b}
\cos(xdm/\hbar t+\phi)$
where $n_a=n_0({\bf r}+{\bf d}/2,t)$ and $n_b=n_0({\bf r}-{\bf d}/2,t)$. 
The resulting fringes are  straight lines orthogonal to the $x$-axis
and have  wave length   
$\lambda = 2\pi\hbar t / md$.
Such  fringes have been directly observed in the experiment of \cite{mit}.

With respect to the experiment of \cite{mit}, based on measurements of the density
of two overlapping condensates, the measurement of the dynamic structure factor 
at high momentum transfer should exhibit  interference even
if the condensates are separated in space.
This  raises the question of what happens
to $S(q,E)$ if the two condensates have never "seen" one another, 
i.e. if they are not in a relatively 
 coherent state before
measurement. According to quantum measurement theory, the system, after 
measurement, jumps into an eigenstate of the 
measured observable. So it is important to understand what are the eigenvalues
of the momentum density operator which is the physical observable
measured in a single realization of the experiment. This should be 
distinguished from the average value taken on several realizations
of the experiment.

According to the Bogoliubov prescription, a fully Bose-Einstein condensed 
state is not only eigenstate of the field operator ${\hat \Psi}$,
but also of the density as well as of the momentum density operators. 
Of course this  prescription does not apply  to 
configurations exhibiting fragmentation of Bose-Einstein 
condensation, as happens in the case of two independent condensates. 
Furthermore it ignores microscopic fluctuations arising from 
the non-commutativity of  ${\hat \Psi}$ and 
${\hat \Psi}^{\dagger}$.  For the above reasons it is interesting 
to discuss in a more quantitative way what are the conditions of 
applicability of the Bogoliubov prescription. 
For a proper discussion of the problem it is crucial 
to consider macroscopic coarse grained averages of the
physical observables.
This averaging  takes
into account the finite resolution of the experimental apparatus.
On the other hand integrating the signal can be 
crucial in order to produce visible interference patterns. From the
theoretical side one can show that only by taking these averages 
will the 
fluctuations of the physical observables 
become negligible.
Let us  discuss in details 
the problem of the momentum distribution which is the 
main object of the present work.
A similar discussion can be repeated  for the 
fluctuations of the density 
operator in the context of experiments where two expanding condensates overlap
in coordinate space and are then imaged \cite{mit}.
Since the momentum distribution  enters the relevant
expression (\ref{Sq}) integrated with respect to $p_y$ and $p_z$, one
only needs
to   consider the 
 coarse grained average along the $x$ direction
\begin{equation}
{\hat \nu}_\beta({p_x})= {1\over \beta\pi^{1/2}}
\int d{\bf p}^{\prime}{\hat n}({\bf p}^{\prime})
\exp[-(p_x-p_x^{\prime})^2/\beta^2] 
\label{coarsep}
\end{equation}
where 
${\hat n}({\bf p}) = {\hat \Psi}^{\dagger}({\bf
p}){\hat \Psi}({\bf p})$
 is the momentum distribution operator and, for simplicity, we 
have chosen a gaussian convolution. 
The fluctuations of the operator 
${\hat \nu}_\beta({p_x})$ are easily calculated for 
a coherent configuration. After  ordering the field operators
one  finds the result
$$
<{\hat \nu}_\beta({p_x}){\hat \nu}_\beta({p_x})> -<{\hat \nu}_\beta({p_x})>^2 = 
{1\over \sqrt{2\pi}\beta} <{\hat \nu}_{\beta/\sqrt2}({p_x})>
$$
where the term in the right hand side arises from the non 
commutativity of the field operators ${\hat \Psi}$ and ${\hat \Psi}^{\dagger}$.
This equation 
explicitly shows that the fluctuations diverge 
when   $\beta \to 0$. Viceversa they become negligible 
in the macroscopic limit $\beta \nu(p_x) \gg 1$, i.e. if 
the number of atoms with momentum between $p_x$ and 
$p_x+\beta$ is sufficiently large. This condition is easily satisfied for
actual
Bose-Einstein condensates and is compatible with the request that 
 $\beta$ be smaller
than the  
distance $\Delta p_x = 2\pi \hbar/d$ between two consecutive fringes
of the momentum distribution. As a consequence of the strong quenching of
the fluctuations, the coherent state can be 
considered, with proper accuracy,  an 
eigenstate of the operator ${\hat \nu}_{\beta}(p_x)$. The
situation is very different if one  instead 
considers two independent condensates occupying, respectively, 
the single particle   wave functions 
$\Psi_a$ and $\Psi_b$. In fact in this case 
the randomness of the
relative phase  produces the additional  contribution 
$<{\hat \nu}_{\beta}(p_x)>^2/2$ to
the fluctuations of  ${\hat \nu}_\beta({p_x})$ 
which become macroscopically large.
 
The above discussion  reveals  that a
state made of two independent condensates  is not 
an eigenstate of the macroscopic observable (\ref{coarsep}). 
As a consequence of  measurement, the system will jump into
a coherent configuration and  will exhibit 
interference patterns. 
Furthermore, if the experiment is non destructive the two condensates
will remain trapped and separated in space also after the measurement.
The possibility of observing interference in a single realization 
of the experiment 
is a peculiar consequence of Bose-Einstein condensation. Of course one should carry out the measurement
within  times shorter than the phase  decoherence times \cite{wallis,castin}.
Different realizations of the experiment on independent condensates
 would give rise to different
values of the phase and consequently to strong fluctuations in the measured
signal.

In conclusion we have pointed out the occurrence of interference phenomena   
in momentum space exhibited  by Bose-Einstein 
  condensates separated in coordinate space. 
The interference patterns should be visible in
the dynamic structure factor measured in  photon scattering experiments
at high momentum transfers.
Such experiments  
 would open 
new intriguing perspectives in the study of 
 quantum superposition and non locality phenomena in macroscopic systems.

Useful discussions with W. Ketterle, A. Leggett,
R. Onofrio and D.M. Stamper-Kurn
are acknowledged.
 This project was  supported by
INFM through the Advanced
Research Project on BEC and by MURST.

%Figure Captions.

%Fig.1. Typical configuration of two separated Bose-Einstein condensates,
%confined by  parallel cigar shaped traps.
 
%Fig.2.
%Dynamic structure factor calculated with (full line) and without 
%(dashed line) coherence between the two condensates. 
%The dynamic structure factor  was calculated in impulse
%approximation, using the Thomas-Fermi limit for the ground
%state momentum distribution (see text). The relative phase 
%was taken  equal to zero, and $\nu =E/2\pi\hbar$.

\begin{figure}
\begin{center}
\epsfig{file=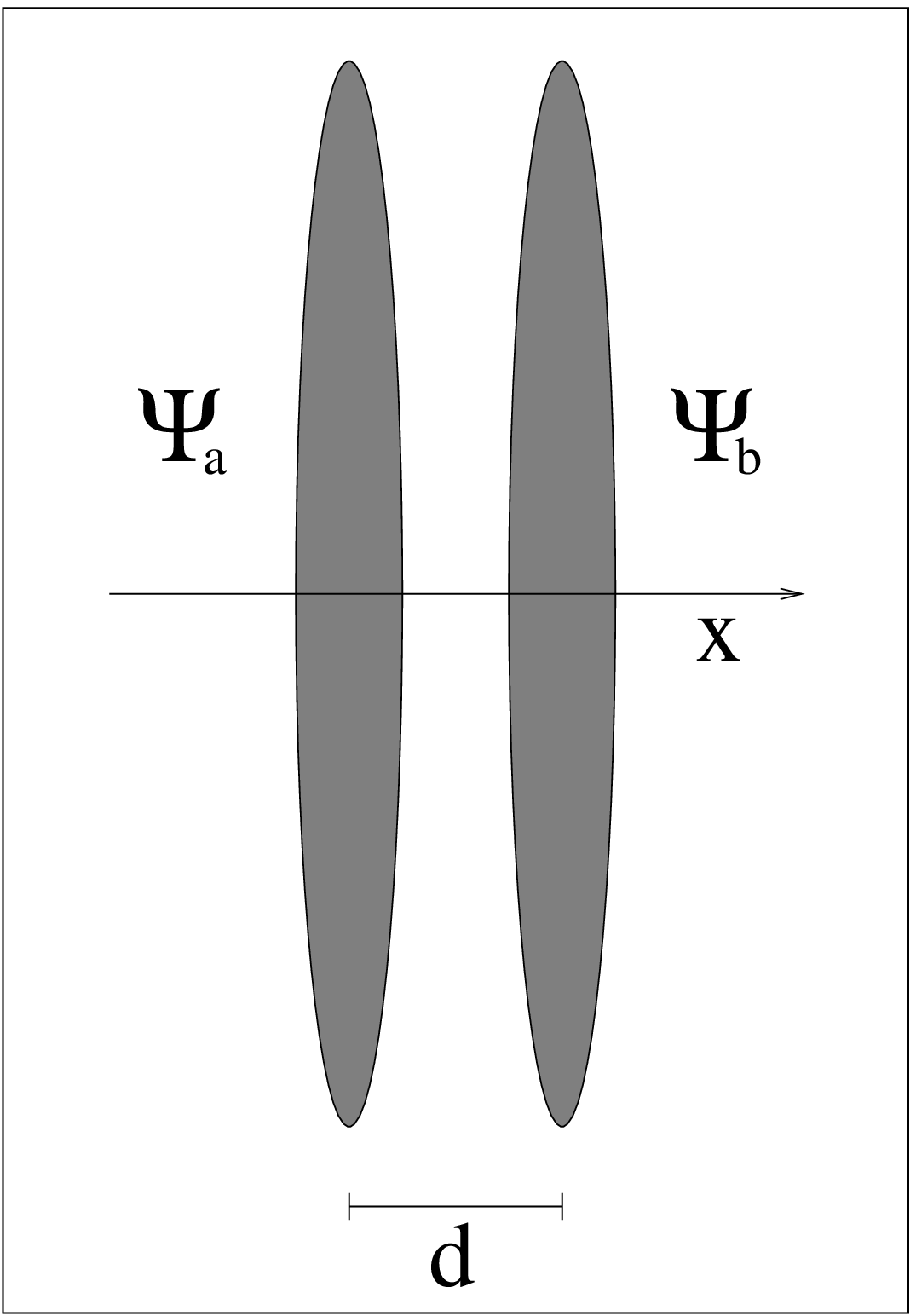, width=0.75\linewidth}
\begin{caption}
{Typical configuration of two separated Bose-Einstein condensates,
confined by  parallel cigar shaped traps.}
\end{caption}
\label{condensati}
\end{center}
\end{figure}

\begin{figure}
\begin{center}
\epsfig{file=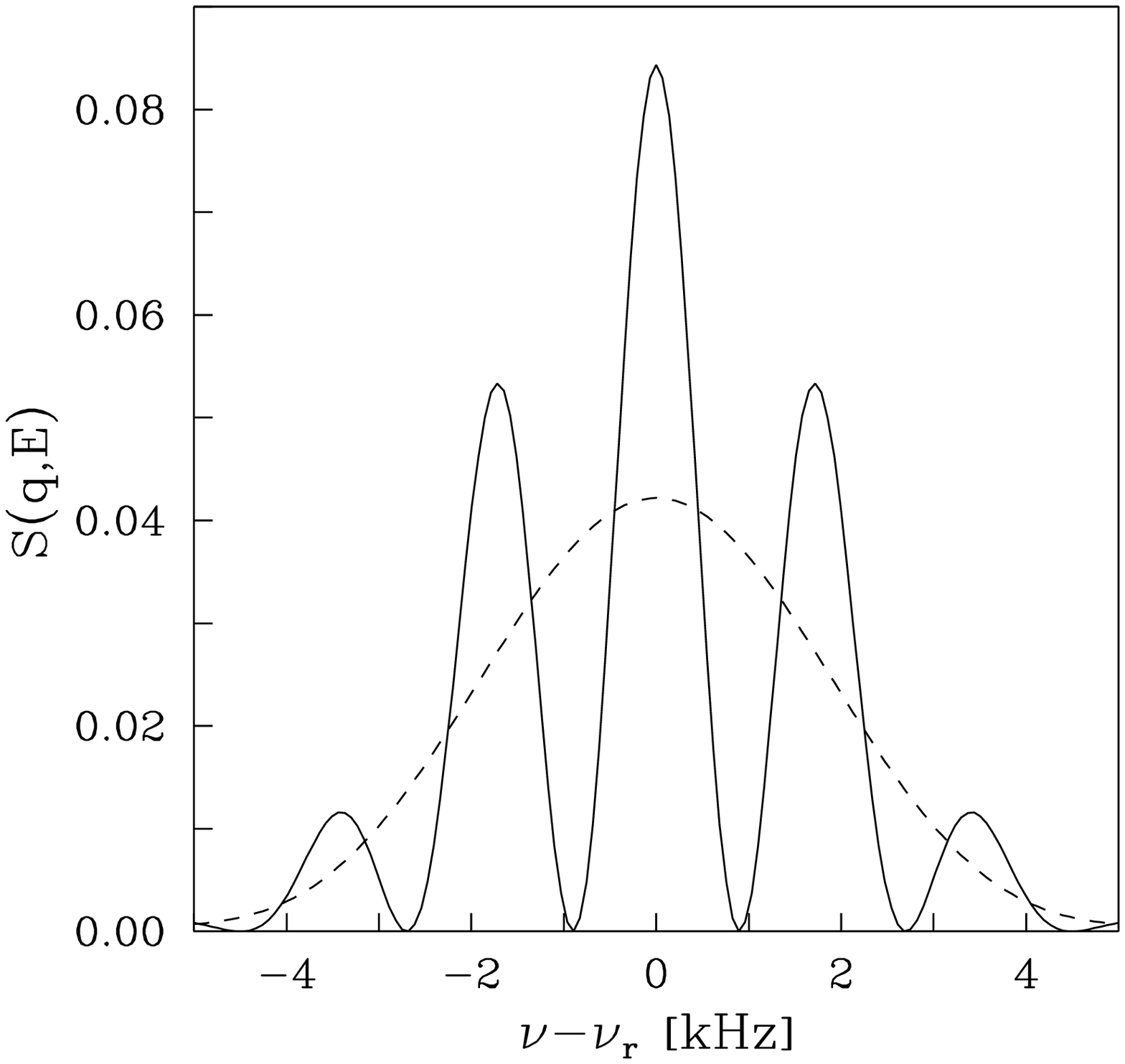, width=\linewidth}
\begin{caption}
{Dynamic structure factor calculated with (full line) and without 
(dashed line) coherence between the two condensates. 
The dynamic structure factor  was calculated in impulse
approximation, using the Thomas-Fermi limit for the ground
state momentum distribution (see text). The relative phase 
was taken  equal to zero, and $\nu =E/2\pi\hbar$. }
\end{caption}
\label{Sqo}
\end{center}
\end{figure}


\begin{references}
\bibitem{note2} This condition is not required if one switches off the
trap immediately after the photon pulse like in \cite{bragg}.
\end{references}
\end{document}